# QUANTUM SEQUENCE STATES

F. V. Mendes,    R. V. Ramos

*Lab. of Quantum Information Technology, Department of Teleinformatic Engineering – Federal University of Ceara - DETI/UFC, C.P. 6007 – Campus do Pici - 60455-970 Fortaleza-Ce, Brazil.*

In a recent paper it has been shown how to create a quantum state related to the prime number sequence using Grover's algorithm. Moreover, its multiqubit entanglement was analyzed. In the present work, we compare the multiqubit entanglement of several quantum sequence states as well we study the feasibility of producing such states using Grover's algorithm.

## 1. Introduction

In the mid-2000s some works connecting quantum information and number theory were reported [1-3] and, more recently, works showing that quantum information is a fertile environment to develop and testing number theory theorems were published [4-6]. In particular, in [5] and [6] Sierra and Latorre showed how to construct the quantum prime state, a quantum sequence state based on the sequence of prime numbers, using Grover's quantum search algorithm. They also studied its multiqubit entanglement. From the best of our knowledge, this is the unique quantum sequence state studied up to now. In this work we consider several different quantum sequence states. We make comparison between their entanglements and study the feasibility of their generation using Grover's quantum algorithm. Furthermore, we introduce a new sequence of integer numbers for which the related quantum sequence state has an entanglement that changes the sign of its slope every time a new qubit is added showing, hence, an oscillatory behavior with period of one qubit.

This paper is outlined as follows. The notation and the quantum sequence states used in this work are presented in Section 2. Section 3 brings the analysis of the entanglement of the quantum sequence states considered. In Section 4, we discuss the feasibility of producing quantum sequence states using Grover's algorithm. At last, Section 5 brings the conclusions.

## 2. Quantum sequence states

From an arbitrary finite integer sequence one can build the related quantum sequence state as follows. Let $S = \{s_1, s_2, s_3, \ldots, s_k\}$ be a set containing the first $k$ elements of an infinite integer sequence, then, a $n$-qubit sequence state related to $S$ with $s_k \leq 2^n-1$, is defined as

$$|S_n\rangle = \left(1/\sqrt{\tau(2^n)}\right)\sum_{i=1}^{k}|s_i\rangle. \qquad (1)$$

In (1) $\tau$ is the sequence counting function of $S$, which returns the sum of the squares of the quantities of each element in the sequence. In the case of a sequence having not repeated elements, the corresponding quantum sequence state is just an equally weighted superposition of the sequence's elements, hence, $\tau(2^n)$ returns the number of elements of $S$ between 0 and $2^n$-1. In this work we will consider the following integer sequences [7]: Fibonacci (A000045), Happy (A007770), Lucky (A000959), Abundant (A005101), Triangular (A000217), Lazy (A000124), Padovan (A000931), Prime (A000040), SPrime (A005097) and PA$^{[r]}$ that is a sequence generated by an arithmetic progression starting from zero and having ration equal to $r$. For example, the four qubit Fibonacci, Happy, Lucky and Prime sequences are

$$|Fib_4\rangle = \frac{1}{\sqrt{10}}(|0000\rangle + 2|0001\rangle + |0010\rangle + |0011\rangle + |0101\rangle + |1000\rangle + |1101\rangle) \qquad (2.a)$$

$$|Hpy_4\rangle = \frac{1}{2}(|0001\rangle + |0111\rangle + |1010\rangle + |1101\rangle) \qquad (2.b)$$

$$|Lck_4\rangle = \frac{1}{\sqrt{6}}(|0001\rangle + |0011\rangle + |0111\rangle + |1001\rangle + |1101\rangle + |1111\rangle) \qquad (2.c)$$

$$|P_4\rangle = \frac{1}{\sqrt{6}}(|0010\rangle + |0011\rangle + |0101\rangle + |0111\rangle + |1011\rangle + |1101\rangle). \qquad (2.d)$$

## 3. Entanglement analysis of sequence states

Since there is not a genuine entanglement measure for a quantum state of arbitrary dimension, in order to analyze the entanglement of the quantum sequence states we will adopt the same strategy used in [5, 8-10]: Given a $n$-qubit quantum state, there are $2^{n-1}$-1 different partial transposes that are relevant to the entanglement measure. The bipartite entanglement of the $i$-th bipartition is given by

$$E_i(|s_n\rangle) = S_{VN}\left[Tr_j(|s_n\rangle\langle s_n|)\right]. \qquad (3)$$

In (3) $S_{VN}$ is the von Neumann entropy, $S_{VN}(\rho)=\text{Tr}[\rho\log(\rho)]$, and $j$ represents the set of qubits traced out. The total amount of entanglement is simply given by the sum of the bipartite entanglement of all relevant bipartitions

$$E_{sum}(|s_n\rangle) = \sum_{i=1}^{2^{n-1}-1} E_i(|s_n\rangle) \tag{4}$$

or its average value

$$E_{avg}^{all}(|s_n\rangle) = \frac{E_{sum}(|s_n\rangle)}{2^{n-1}-1}. \tag{5}$$

However, the calculation of (5) requires a considerable computational effort when the number of qubits grows. An alternative and easier to calculate entanglement measure is the average between the entanglement of bipartitions formed by one qubit and $n$-1 qubits (hence, only $n$ bipartitions are used). We use the upper level index to indicate these particular bipartitions. Thus, for an $n$-qubit state one has

$$E^1(|s_n\rangle) = S_{VN}\left[Tr_1(|s_n\rangle\langle s_n|)\right], E^2(|s_n\rangle) = S_{VN}\left[Tr_2(|s_n\rangle\langle s_n|)\right],\ldots,E^n(|s_n\rangle) = S_{VN}\left[Tr_n(|s_n\rangle\langle s_n|)\right] \tag{6}$$

$$E_{avg}^{th}(|s_n\rangle) = \frac{1}{n}\sum_{i=1}^{n} E^i(|s_n\rangle). \tag{7}$$

Figure 1 shows $E^i$ for six sequence states of 28 qubits.

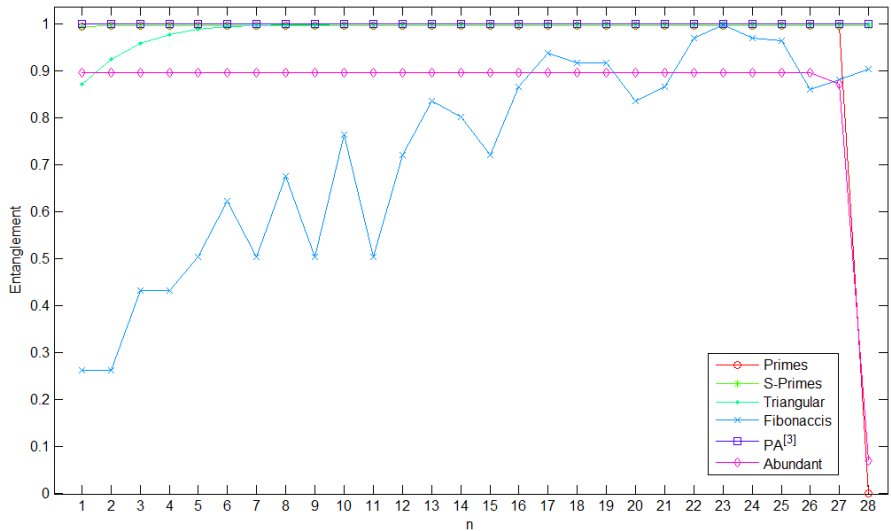

Figure 1: $E^i$ versus $i$, for Prime, S-prime, Triangular, Fibonacci, PA[3] and abudant sequences.

Observing Fig.1, one can note that the entanglement of each individual qubit with the others 27 qubits of the Abundant and Prime states has a similar behavior: with exception of the last qubit that has low entanglement with the others (zero in the case of the Prime state), all the others are highly entangled with the rest of the state. The explanation given in [5] associates this behavior with the fact that the Prime sequence is formed almost exclusively by odd numbers. This explanation cannot be used for the Abundant state that has a significant amount of both, odd and even numbers. The Fibonacci state, by its turn, shows a very different behavior, there is a low entanglement between the first qubits and the rest, furthermore, the value of $E^i$ changes in a non-regular way when $i$ grows. A resume of average entanglement values given by (7) is shown in Table 1.

Table 1: $E_{avg}^{th}$ for 28-qubits quantum sequence states.

| State | $E_{avg}^{th}$ | State | $E_{avg}^{th}$ |
|---|---|---|---|
| Prime | 0.9606 | Triangular | 0.9897 |
| SPrime | 0.9964 | Abundant | 0.8660 |
| Fibonacci | 0.7302 | PA[3] | 1.0000 |

The entanglement of the quantum sequences SPrime, Triangular, Abundant and PA[3] overcomes the entanglement of the Prime state. In particular, the sequence PA[3] shows the maximal entanglement value, hence, each individual qubit of PA[3] is in the maximally mixed state. The relation between the entanglement of the series changes when the average between all bipartitions is taken into account. Figure 2 shows the comparison of $E_{avg}^{th}$ and $E_{avg}^{all}$ between PA[3] and Fibonacci states.

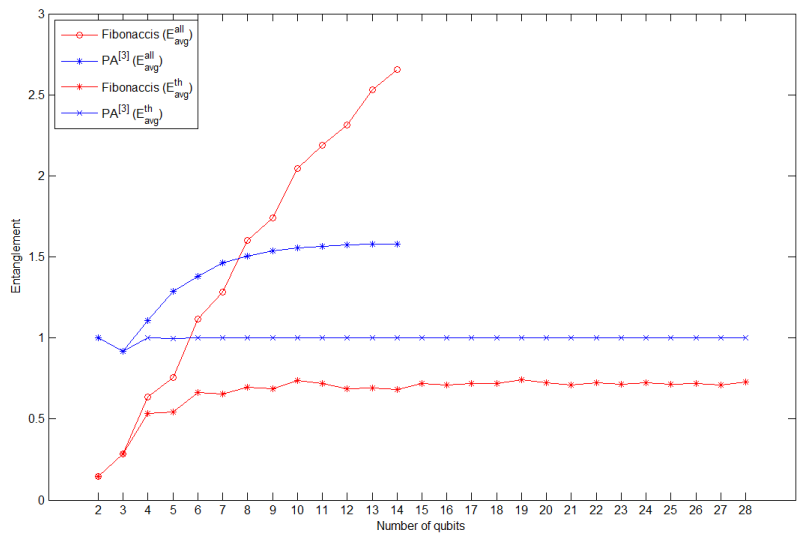

Figure 2: Comparison of $E_{avg}^{th}$ and $E_{avg}^{all}$ between Fibonacci and PA[3] states.

As it can be seen in Fig. 2, one has $E_{avg}^{th}(|PA_n^{[3]}\rangle) > E_{avg}^{th}(|Fib_n\rangle)$ for any number of qubits up to 28, while $E_{avg}^{all}(|PA_n^{[3]}\rangle) < E_{avg}^{all}(|Fib_n\rangle)$ for sequence states having more than six qubits.

It is not an easy task to understand for which reason a sequence state shows a large amount of entanglement. A supposition made in [5] is that the entanglement of the Prime state emerges from intrinsic randomness of prime numbers. However, as it can be seen in Fig. 1 and Tab. 1, using $E_{avg}^{th}$ as reference, there are quantum sequence states originated from deterministic sequences whose entanglement is larger than the entanglement of the Prime state. The more regular of them is the $PA^{[r]}$ state. In Fig. 3 one can see the entanglements $E_{avg}^{th}$ and $E_{avg}^{all}$ of $PA^{[r]}$ for $r = 3, 5, 7, 9$.

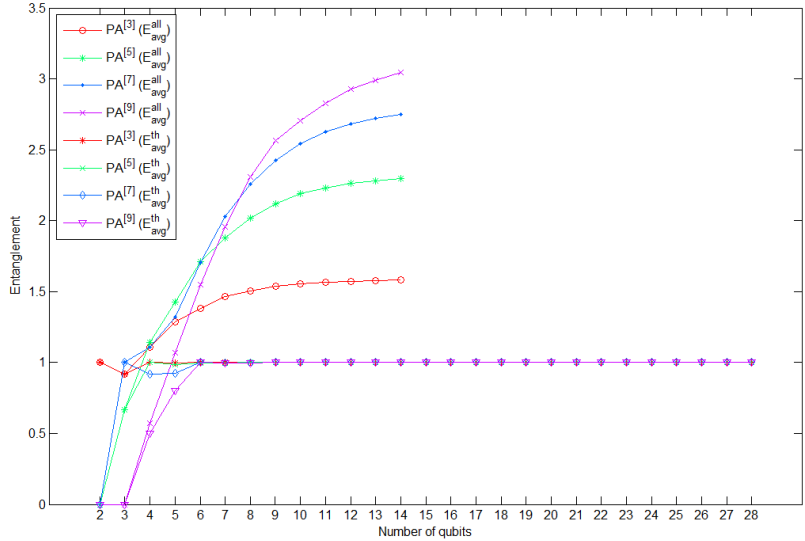

Figure 3: $E_{avg}^{th}$ (up to 28 qubits) and $E_{avg}^{all}$ (14 qubits) for $PA^{[r]}$, $r = 3, 5, 7, 9$.

Observing Fig. 3, one can note that the $PA^{[r]}$ state tends to have most qubits maximally entangled with the others. Furthermore, for the set of values of $r$ considered, after the seventh qubit, the larger the value of $r$ the larger is $E_{avg}^{all}$. On the other hand, it is not hard to see that $PA^{[r]}$ states do not have any entanglement when $r$ is a power of two. This fact can be seen in Fig. 4 that shows $E_{avg}^{all}(|PA_{14}^{[r]}\rangle)$ versus $r$ for a quantum state with 14 qubits. In fact, for $r = 2^k$ one has the following (completely disentangled) sequence state:

$$\left|PA_n^{[r=2^k]}\right\rangle = \begin{cases} |0\rangle^{\otimes n} & n < k \\ (H|0\rangle)^{\otimes(n-k)} \otimes |0\rangle^{\otimes k} & n \geq k \end{cases}. \tag{8}$$

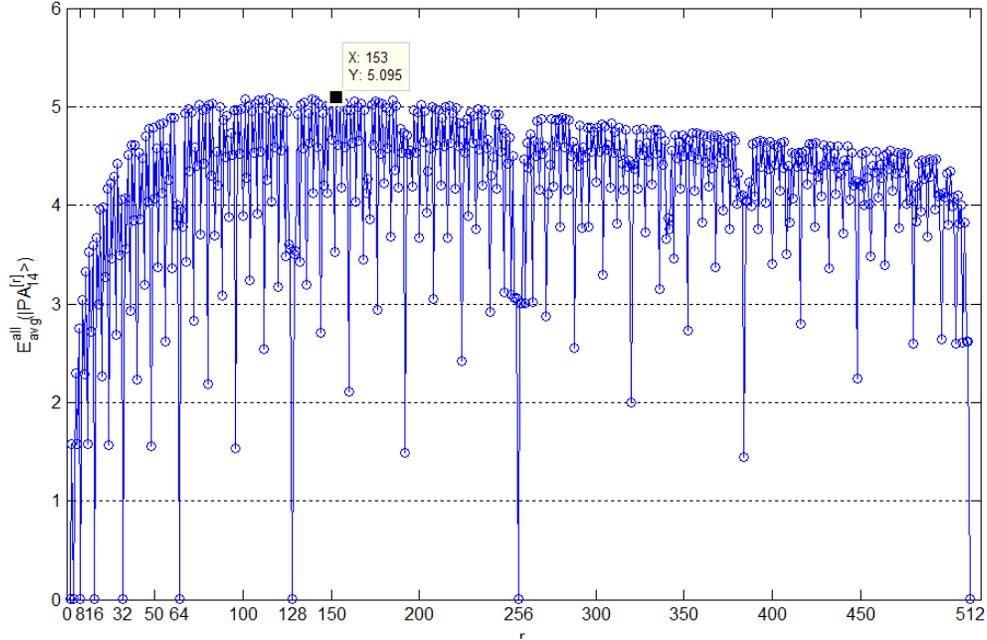

Figure 4: $E_{avg}^{all}\left(\left|PA_{14}^{[r]}\right\rangle\right)$ verus $r$.

In [5,6] is pointed out that the Prime state carries a large amount of entanglement but, in fact, how near from a maximally entangled (using (4)) $n$-qubit state is the $n$-qubit Prime state? Figure 5 shows $E_{avg}^{all}$ for five different quantum sequence states.

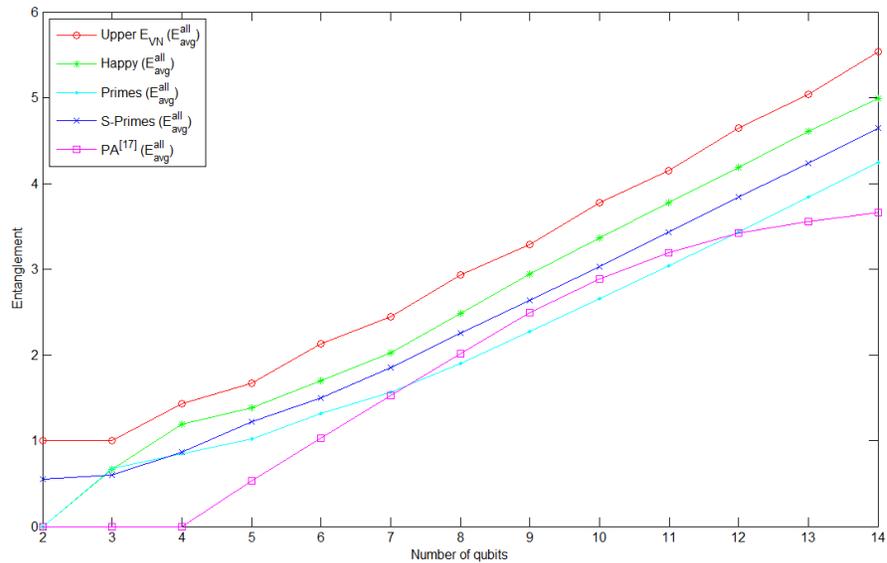

Figure 5: $E_{avg}^{all}$ versus number of qubits and its comparison with the maximal value reacheable by the measure. Happy, Prime, Sprime and PA[17] sequences.

Analyzing Fig. 5, one can see that, inside the short search space observed, the entanglement of the states SPrime and Happy overcomes the Prime state's entanglement. While the Prime state reaches nearly 68% of predicted upper bound, the SPrime and Happy states reach, respectively, 74% and 78%. On the other hand, the entanglement of the PA[17] state overcome the Prime state's entanglement only in a short range, after, it begins to move away from the upper bound. In Fig. 6 other sequence states are shown.

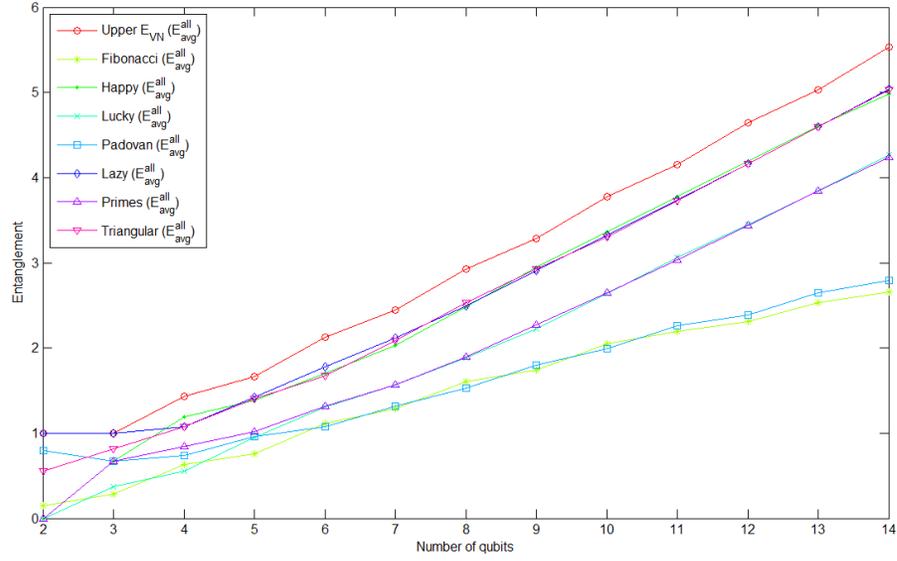

Figure 6: $E_{avg}^{all}$ versus number of qubits and its comparison with the maximal value reacheable by the measure. Fibonacci, Happy, Lucky, Padovan, Lazy, Primes and Triangular sequences.

The Fibonacci and Padovan states follow a similar behavior and they have only five common elements (~7%). Similarly, the Lucky and Prime states follow a similar behavior and they have only ~7% of common elements. Furthermore, the Happy, Lazy and Triangular states also have a common entanglement behavior.

Now, let us introduce a new sequence, hereafter named $S$. In order to construct $S$ we start with the sequence $S = \{0,3\}$. The sequence $S$ is obtained iteratively according to the following steps ($k = 2,3,4,…, n$-1):

- If $k$ is even, then $S = S \cup \max(\{1,2,…,2^k\text{-}1\}\text{-}S)$ and $k = k+1$. This step adds only one number to the sequence.

- If $k$ is odd, then $S = S \cup bitxor(S,2^{k+1}\text{-}1)$ and $k = k+1$. If $X$ is the set $\{x_1,x_2,x_3,…,x_n\}$ and $Y$ is just a number, then $bitxor(X,Y)$ is the set $\{Dec(Bin(x_1)\oplus Bin(Y)), Dec(Bin(x_2)\oplus Bin(Y)),…,$

Dec(Bin($x_n$)⊕Bin($Y$))}. Here Dec and Bin are functions that return, respectively, the decimal and binary value of the argument.

For example, let us construct the sequence that finishes at $k = 6$.

| | |
|---|---|
| $k=2$ | $S = \{0, 3\} \cup \max(\{1, 2, 3\} - \{0, 3\}) \rightarrow S = \{0, 3\} \cup \max(\{1, 2\}) \rightarrow S = \{0, 3\} \cup \{2\} \rightarrow S = \{0, 2, 3\}$. $k = 2 + 1$. |
| $k=3$ | $S = \{0, 2, 3\} \cup \text{bitxor}(\{0, 2, 3\}, 15) \rightarrow S = \{0, 2, 3, 12, 13, 15\}$. $k = 3 + 1$. bitxor({0, 2, 3}, 15) = {(0⊕15), (2⊕15), (3⊕15)} = {15, 13, 12} |
| $k=4$ | $S = \{0, 2, 3, 12, 13, 15\} \cup \max(\{1, 2, 3, \ldots, 15\} - \{0, 2, 3, 12, 13, 15\}) \rightarrow S = \{0, 2, 3, 12, 13, 15\} \cup \max(\{1, 4, \ldots, 11, 14\}) \rightarrow S = \{0, 2, 3, 12, 13, 15\} \cup \{14\} \rightarrow S = \{0, 2, 3, 12, 13, 14, 15\}$. $k = 4 + 1$. |
| $k=5$ | $S = \{0, 2, 3, 12, 13, 14, 15\} \cup \text{bitxor}(\{0, 2, 3, 12, 13, 14, 15\}, 63) \rightarrow S = \{0, 2, 3, 12, 13, 14, 15\} \cup \{48, 49, 50, 51, 60, 61, 63\} \rightarrow S = \{0, 2, 3, 12, 13, 14, 15, 48, 49, 50, 51, 60, 61, 63\}$. $k = 5+1$. bitxor{0,2,3,12,13,14,15},63) = {(0⊕63), (2⊕63), (3⊕63), (12⊕63), (13⊕63), (14⊕63), (15⊕63)} = {48, 49, 50, 51, 60, 61, 63} |
| $k=6$ | The procedure ends. |

Hence, the sequence generated in this example is $S = \{0,2,3,12,13,14,15,48,49,50,51,60,61, 63\}$. The quantum sequence state $|S\rangle$ has an interesting behavior, its entanglement shows an oscillation with period of one qubit. In Figs. 7 and 8 one can see, respectively, the oscillatory behavior of $E_{avg}^{th}$ for $|S_{28}\rangle$ and $E_{avg}^{all}$ for $|S_{14}\rangle$.

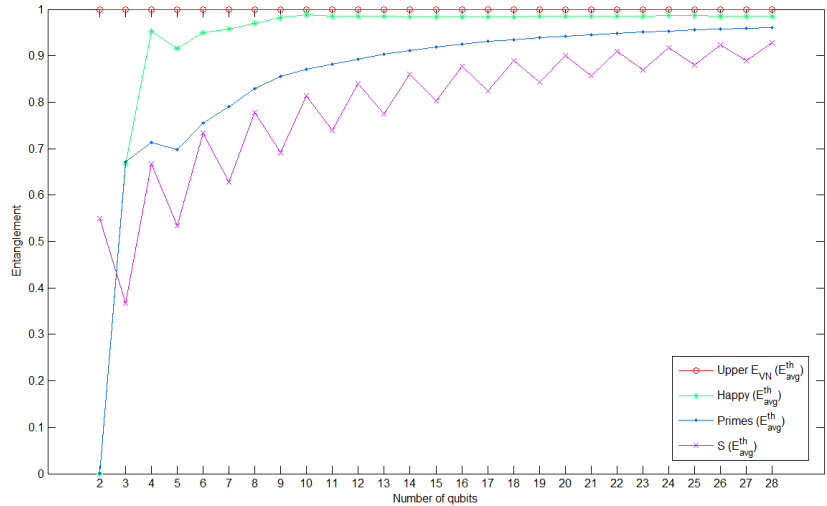

Figure 7: $E_{avg}^{th}$ versus number of qubits for $|S_{28}\rangle$.

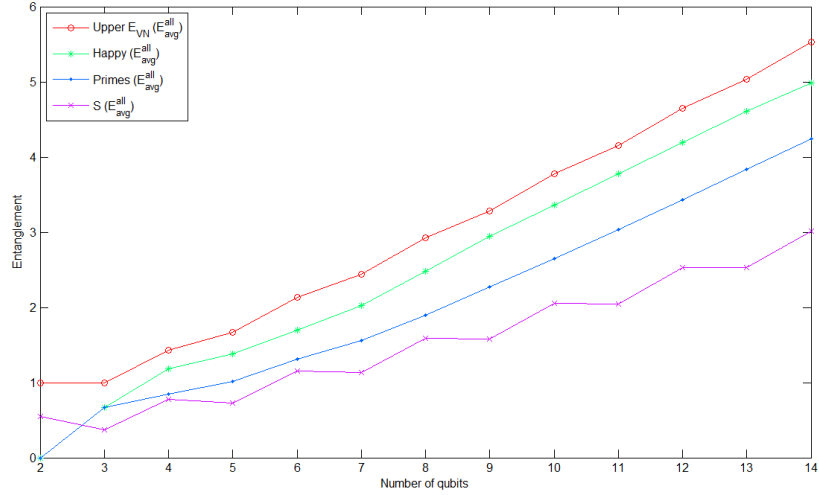

Figure 8: $E_{avg}^{all}$ versus number of qubits for $|S_{14}\rangle$.

## 4. Quantum sequence state preparation with Grover's quantum search

The approach proposed to prepare the Prime state in [5] can also be used to prepare other quantum sequence states, provided that there is an oracle able to check whether a given element belongs to the considered sequence. Basically, the algorithm searches for $\tau(2^n)$ items within a set of $2^n$ elements. Grover's algorithm accomplishes this in $O((\tau(2^n)/2^n)^{1/2})$. The optimal value of iterations is given by

$$G(n) = \frac{\pi}{4\arcsin\left(\sqrt{\frac{\tau(2^n)}{2^n}}\right)} - \frac{1}{2}. \qquad 9$$

Hence, the feasibility of the sequence state generation using quantum search depends on how $G(n)$ grows when $n$ (the number of qubits) increases. Figure 9 shows the curve of $G(n)$ versus $n$ for Fibonacci, Lucky, Padovan, Lazy and Triangular quantum sequence states.

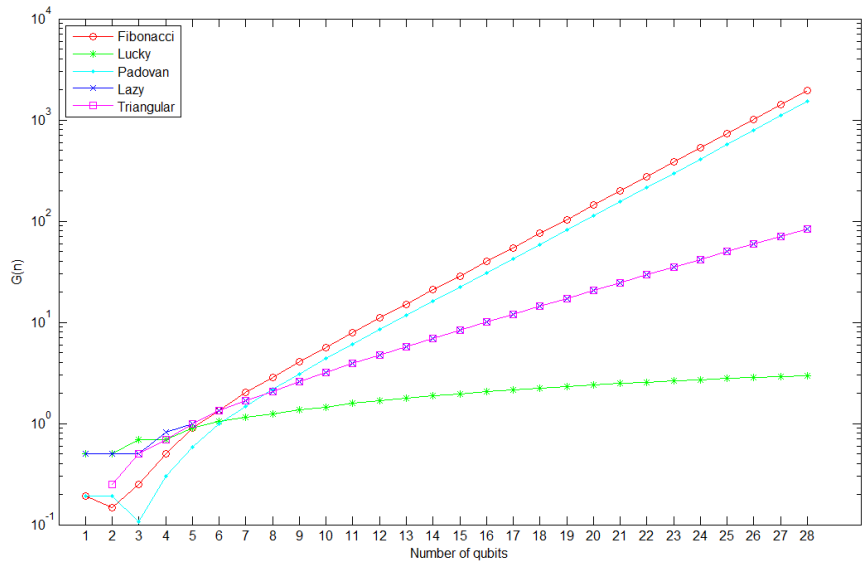

Figure 9: Number of Grover's iterations versus number of qubits for Fibonacci, Lucky, Padovan, Lazy and Triangular quantum sequence states.

Figure 10 shows a similar plot for Abundant, Happy, Hashard, Lucky, Prime and SPrime sequence states. The Abundant and Happy sequence states seem to assume a constant behavior ($\tau(2^n)/2^n$ remains roughly constant). The efficiency of generation of the SPrime state overcomes the efficiency of generation of the Prime state.

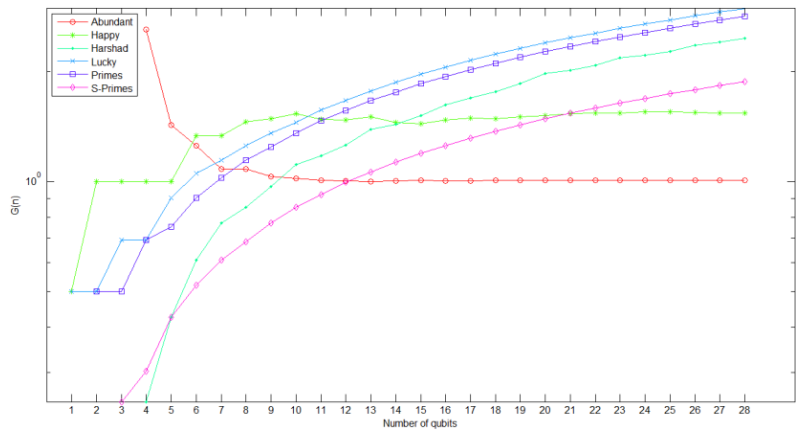

Figure 10: Number of Grover's iterations versus number of qubits for Abundant, Happy, Hashard, Lucky, Prime and SPrime sequence states.

Regarding the PA states, after a growing initial part, a constant behavior appears, as can be seen in Fig. 11.

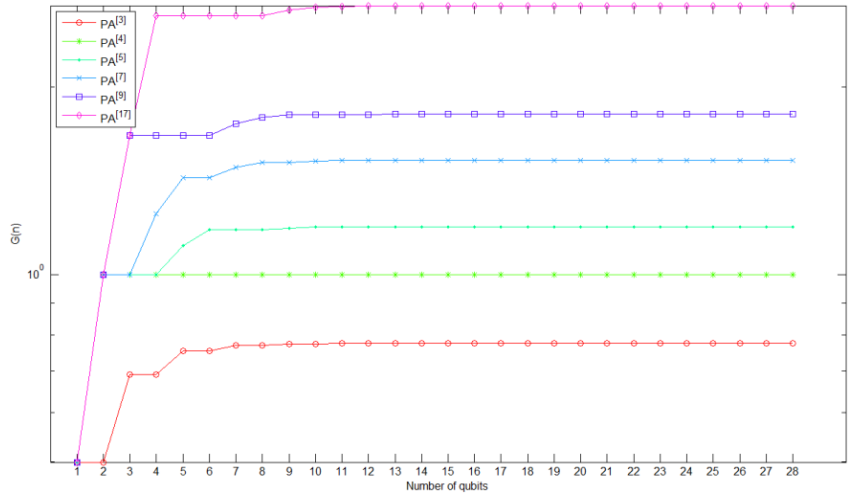

Figure 11: Number of Grover's iterations versus number of qubits for PA$^{[r]}$ sequences, $r \in \{3,4,5,7,9,17\}$.

At last, the curve of $G(n)$ for the state $|S\rangle$ with oscillatory entanglement can be seen in Fig. 12.

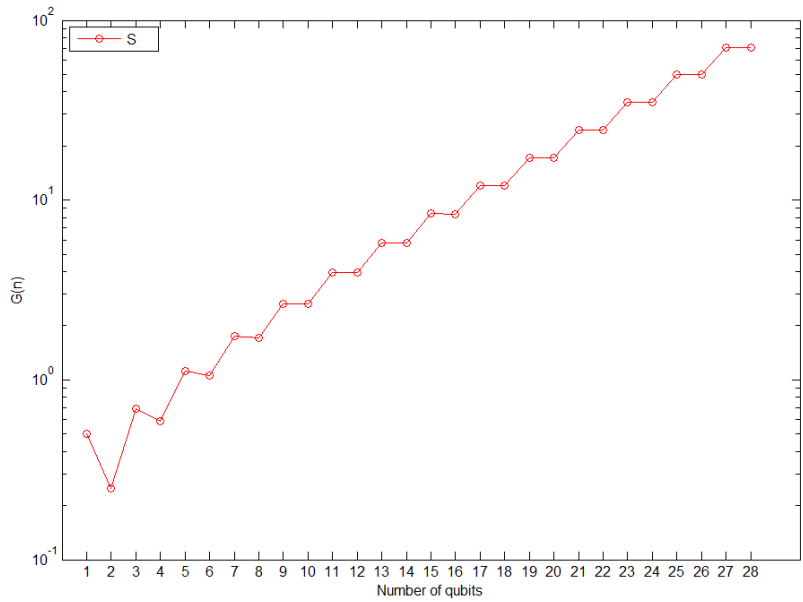

Figure 12: Number of Grover's iterations versus number of qubits for the sequence state $|S\rangle$.

## 5. Conclusions

The analysis of the entanglement of sequence states is a hard task. Sequences states with similar entanglement behavior may have a common pattern but such pattern is not readily observed just looking at their elements. Prime and Lucky states are good examples, as well Fibonacci and Padovan states.

The Prime state has the charm of being related to prime numbers that play a crucial role in number theory. However, the Prime state is not the most entangled sequence state (for example, the Happy and SPrime sequences have more entanglement) as well it is not efficiently produced by quantum search.

Trying to connect entanglement with a pseudo-randomness of the numbers that make-up the sequence seems not to be a fruitful path since there are sequence states that emerge from a trivial increment pattern, which can carry a significant amount of entanglement. In particular, the sequence state PA[3] has maximal entanglement between each individual qubit and the rest $n$-1 qubits.

The way in which the entanglement changes when a qubit is added depends on the sequence considered. All sequences obtained from [7] showed a (growing) smooth behavior of $E_{avg}^{all}$. However, this is not a rule. In order to show this, we created a sequence whose related quantum state shows an (growing) oscillatory behavior of $E_{avg}^{all}$.

The feasibility of sequence state preparation using Grover's algorithm depends on the relation $\tau(2^n)/2^n$: if the number of integers that belong to the series grows in a rate lower than 2 when a qubit is added, then the use of Grover's algorithm is not viable for large values of $n$. This happens for Fibonacci, Lucky, Lazy, Padovan, Triangular, Sprime, Prime and the introduced $S$ series. On the other hand, the Abundant, Happy and PA series can be efficiently produced with Grover's algorithm.

## Acknowledgments

This work was supported by the Brazilian agency CNPq Grant no. 303514/2008-6. Also, this work was performed as part of the Brazilian National Institute of Science and Technology for Quantum Information.